# Grid-Based Multichannel Access in Vehicular Networks

Yalew Zelalem Jembre and Young-June Choi
Department of Computer Engineering
Ajou University, Suwon 443-749, South Korea
Email: {zizutg, choiyj}@ajou.ac.kr

*Abstract*— In vehicular networks, vehicles exchange messages with each other as well as infrastructure to prevent accidents or enhance driver's and passenger's experience. In this paper, we propose a grid-based multichannel access scheme to enhance the performance of a vehicular network. To determine the feasibility of our scheme, we obtained preliminary results using the OPNET simulation tool.

*Keyword*s— Vehicular network, V2V, V2X, channel assignment, multichannel access

## I. Introduction

Wireless vehicular communication was initially conceived to be a network of nearby vehicles to exchange warning messages to prevent collisions as well as possible blind spots, which are mostly safety applications. However, due to recent advance of devices and wireless technology, other applications such as media download or parking management (generally referred to as non-safety applications) are now included as part of services provided by vehicular networks. Messages are sent from a vehicle to everything (V2X) including vehicle-to-vehicle (V2V) and vehicle-to-infrastructure (V2I). Here, infrastructure usually refers to road side units (RSUs) where RSUs are similar to access points that are intended to connect vehicles to the Internet or other infrastructure [1].

Although multiple standardizations are currently on the table [2], this work, focuses on Dedicated Short Range Communications (DSRC) [3] that is currently being deployed. The Federal Communication Commission (FCC) of the United States has allocated 75 MHz of bandwidth in 5GHz spectrum for vehicular communication, which are divided into 7 channels (172-184), each of which has 10 MHz of bandwidth. In DSRC the IEEE 1609.4 a protocol that lies above 802.11p, is responsible for the multichannel operation. IEEE 802.11p is the amendment of IEEE802.11a DCF MAC. IEEE 1609.4 uses one channel (178) as a control channel (CCH) and the rest as service channels (SCHs). Vehicles in DSRC are synchronized every 100ms through a coordinated universal time (UTC), which is divided into two intervals; CCH and SCH interval. During CCH interval every vehicle switches to CCH (178) and receives safety information as well as other control information whereas in SCH interval vehicles tune to least congested SCHs and communicate non-safety applications [4]. IEEE 1609.4 is mostly designed with a single interface (SI) in mind, while multi interface (MI) is considered for future V2X networks. Despite the importance, CCH is easily prone to congestion [2].

Multichannel (MC) operation in V2X still remains an open issue as pointed out in [1-4]. The recent work found in [5],

This research was supported by Basic Science Research Program through the National Research Foundation of Korea (NRF) funded by the Ministry of Education (2015R1D1A1A01058025).

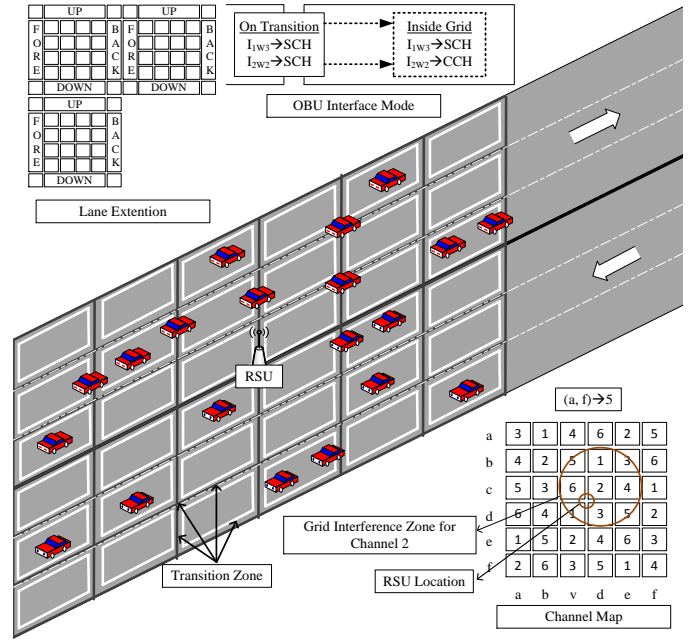

Figure 1 Proposed multichannel access architecture

which is Time Division Multiple Access (TDMA) based MAC, is a good example that addressed multichannel operations with dual radio while guaranteeing a transmission for every vehicle within a reasonable time. This work assigns one interface to CCH and the others to SCH. However, MI can be used to enhance vehicular communication even more, if a better access scheme is used. Motivated by this, we propose a new MI-based MC operation.

## II. Proposed model

Due to fast speed in vehicular networks, MC access with a dynamic assignment as well as switching is not suitable. This is because vehicles will either accelerate or decelerate before even using the channel, which will render channels unused. Vehicles have a predetermined area and trajectory that can be used as an advantage. Therefore, we propose a grid-assisted MC operation that suits the deployment area with a predetermined channel map.

In our model, each vehicle's on-board unit (OBU) is equipped with two interfaces ($I_1$ and $I_2$), configured in W3 and W2 modes respectively; W3 mode can be tuned to SCHs only whereas W2 mode can be tuned to both CCH and SCHs [2]. First we divide the road into a 6x6 matrix, because there are six channels as shown in Figure 1. At least one RSU is placed at the center. When a vehicle turns on for the first time or joins the network it receives its grid's coordinates and distance from the four transition zones (TZs) from RSU. TZ is a space located before the end of a grid, which is equal for all sides of a grid, for instance, 10 meters to the edge. However, vehicles are responsible for predicting the next grid based on its channel map, speed and trajectory. Each grid is assigned one SCH. Since the number of lanes varies, to cover the topology of a V2X network we use the lane extension rule, as depicted in Figure 1.

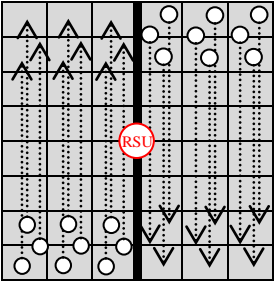

Figure 2 Simulation setup model

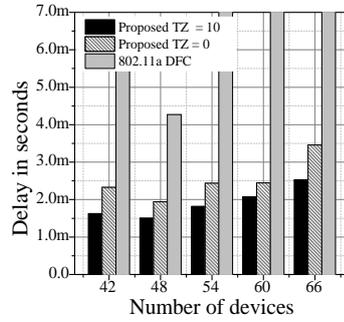

Figure 3 Delay as the number of users increase

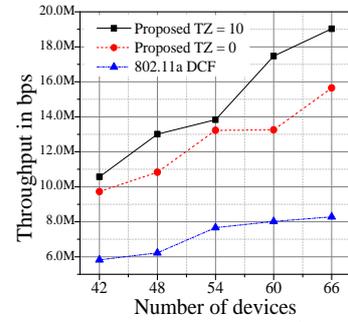

Figure 4 Throughput as the number of users increase

The assignment of the grid is done using N-Queens strategy [6]. The benefit of such a strategy is two-folds, 1) it reduces interference since channels are reused outside grids interference zone (GIZ) and 2) guarantees fairness because every channel is proportionally assigned and accessed uniformly. Interfaces switch channel based on the following rules, i) $I_1$ *is tuned to the SCH of the grid that the vehicle resides in*; ii) $I_2$ *is tuned to CCH unless the vehicle is in TZ*; iii) *when the vehicle is in TZ $I_2$ is assigned to the SCH of the next grid*. This assignment benefits both the vehicle and the network. From vehicles' point of view it provides seamless communication since it will not discontinue communication during channel switching. Meanwhile, vehicles on the TZ access two SCHs, which can be used to relay service between grids to enhance connectivity in the network.

## III. RESULT AND DISCUSSION

We have performed a preliminary simulation using OPNET [7]. In this section, we will discuss the findings and the shortcomings of the proposed model. Since DSRC amends 802.11a and adds the 1609.4 extension for MC operations, we compared the proposed scheme with 802.11a distributed coordination function (DCF) MC operation. Two variants of the proposed scheme are compared by varying the value of TZ i.e. TZ is 10 meters from the edge and 0 meters. We have deployed a 6 columns and 18 rows grid in 1000m$^2$ area while varying the number of nodes. The right and left sides of the road have three columns each, and the number of vehicles is divided in half for each side as shown in Figure 2. The speed and the transmission power of each node are set to 50 km/hr and 0.001w, respectively.

In vehicular networks, especially for safety and control application delivery, delay should be kept as minimum as possible. For that reason, we first compare the delay incurred by each scheme and found that the proposed scheme with TZ of 10m has the lowest delay in all the scenarios. Since 802.11a DCF uses the least congested channel assignment method, it incurs the highest delay switching between channels, to find the least congested, as shown in Figure 3. Throughput of the network, which is mostly relevant to nonsafety applications, is compared in Figure 4, which depicts the result as the number of devices in the network increases. Similar to delay TZ has actually benefited the network. However, the value of TZ requires a deeper investigation or even be made adaptive to the number of nodes in the network.

In this work, we have started our study to find a new MC access scheme for vehicular networks. Based on the characteristics of vehicular networks, we have proposed a scheme that is suitable to the network. To confirm the performance of the scheme we have tested it with initial level network setup and implementation. Although the performance obtained from the simulation encouraged us to proceed with current research direction, there are other metrics that need to be taken into consideration, such as overhead and congestion, to completely confirm its superiority. In addition, it should be compared to other MC schemes that are similar to [6]. The main drawback of this work is adapting to the change in the number of channels. That means, it will accommodate the available number of channels between 6 and 18 to guarantee channel access uniformity. Therefore, we plan to include the following extensions in the future work, 1) improve the scheme to avoid congestion on CCH 2) detailed description of the scheme; 3) extensive simulation with more realistic environment and metrics; 4) further comparison with other existing studies.